\documentclass[11pt]{article}
\usepackage{moriond,epsfig}

\def\be{\begin{equation}}
\def\ee{\end{equation}}
\def\bea{\begin{eqnarray}}
\def\eea{\end{eqnarray}}

\usepackage{epsfig}
\usepackage{graphicx}               
\usepackage{dcolumn}                
\usepackage{multirow}               
\usepackage{amsmath}                
\usepackage{color}
\usepackage{comment}
\graphicspath{{ps}}                 

\newcommand{\mpp}{M_{p\bar{p}}}

\newcommand{\bp}{B^{+}}

\newcommand{\bz}{B^{0}}

\newcommand{\pp}{{p\bar{p}}}
\newcommand{\ppk}{{p\bar{p}K^+}}
\newcommand{\pppi}{{p\bar{p}\pi^+}}

\newcommand{\mll}{M_{\Lambda\bar{\Lambda}}}
\newcommand{\LL}{\Lambda\bar{\Lambda}}
\newcommand{\LcLc}{\Lambda_c^+ \bar{\Lambda}_c^-}

\newcommand{\XcLcm}{\Xi_c^0 \bar{\Lambda}_c^-}
\newcommand{\llk}{{\Lambda\bar{\Lambda}K^+}}
\newcommand{\llpi}{{\Lambda\bar{\Lambda}\pi^+}}
\newcommand{\llkz}{{\Lambda\bar{\Lambda}K^0}}

\newcommand{\llkst}{{\Lambda\bar{\Lambda}K^{* +}}}
\newcommand{\llkstz}{{\Lambda\bar{\Lambda}K^{*0}}}

\begin{document}
\vspace*{4cm}

\title{BARYONIC \textbf{\textit{B}} MESON DECAYS AT BELLE AND BABAR}

\author{ Y. W. Chang }

\address{Department of Physics, National Taiwan University, 
         No. 1, Sec. 4, Roosevelt Road, Taipei, 10617 Taiwan (R.O.C)}

\maketitle\abstracts{
Recent results obtained using the data sample 
collected on the $\Upsilon(4S)$ resonance
with the Belle detector at the KEKB asymmetric-energy $e^+ e^-$ collider and the Babar detector at the PEP-II 
asymmetric-energy $e^+ e^-$ collider are discussed. 
Measurements of several charmless and charmed baryonic $B$ decay branching fractions are reported, and 
some behaviors and mechanisms are discussed.}

\section{Introduction}
Baryonic decays of $B$ mesons, which contain a heavy bottom quark and a light up or down quark, 
can provide a laboratory for a range of particle physics investigations like:
measurements of baryonic decay branching fractions;
trends in decay rates and baryon production mechanisms;
searches for exotic states, 
excited baryon resonances and intermediate states; 
examination of the angular distributions of $B$-meson decay products to determine baryon spins;
and measurements of radiative baryonic B decays that could be sensitive to new physics through flavor-changing neutral currents.
In 2002, a charmless three-body baryonic $B$ decay, $\bp \to \ppk$~\cite{KAbe}, was
observed at Belle. Moreover, the peaking distribution near the threshold
in $\mpp$ spectrum displays that larger energy meson is flavored during the
decay process and matches the theoretical expectation. After this exhortative
discovery, many other similar observations were accomplished, and many theoretical calculations were done for the sake 
understanding the mechanism of these decay modes.
In this paper, we'll report some results of baryonic $B$ decays, studied in 2007 $\sim$ 2009.

\section{Charmless baryonic \textit{B} decays}
\subsection{Observation of $B^0 \to \Lambda \bar{\Lambda} K^0$ and $B^0 \to \Lambda \bar{\Lambda} K^{*0}$}
In this paper~\cite{YWChang}, the charmless three-body decays $B \to \LL h$ are studied, 
where $h$ stands for $\pi^+$, $K^+$, $K^0$, $K^{*+}$, or $K^{*0}$, using a $605 fb^{-1}$ data sample collected 
at the $\Upsilon(4S)$ resonance with the Belle detector at the KEKB asymmetric energy $e^+ e^-$ collider. 
One can simply replace the $sd - \bar{s}\bar{d}$ diquark pair with an $ud - \bar{u}\bar{d}$ pair 
to establish a one-to-one correspondence between $B \to \LL h$ and $B \to \pp h$ decays. 
A common feature of these decays is that the baryon-antibaryon mass spectra peak near threshold.
The $K^+$ meson carries the energetic $\bar{s}$ quark from the $\bar{b} \to \bar{s}$ transition 
so that a threshold enhancement of the baryon-antibaryon system is naturally formed. 
However, there is another possibility that the $\bar{\Lambda}$ (instead of the $K^+$) 
carries the $\bar{s}$ from the $\bar{b} \to \bar{s}$ transition. 
It is interesting to know the role of this $\bar{s}$ quark in $B \to \LL K^{(*)}$ weak decays.

\begin{figure}[tph]
\begin{center}
 \hskip -5.8cm {\bf (a)} \hskip 4.5cm {\bf (b)}\\
 \vskip -0.8cm
  \includegraphics[width=0.33\textwidth]{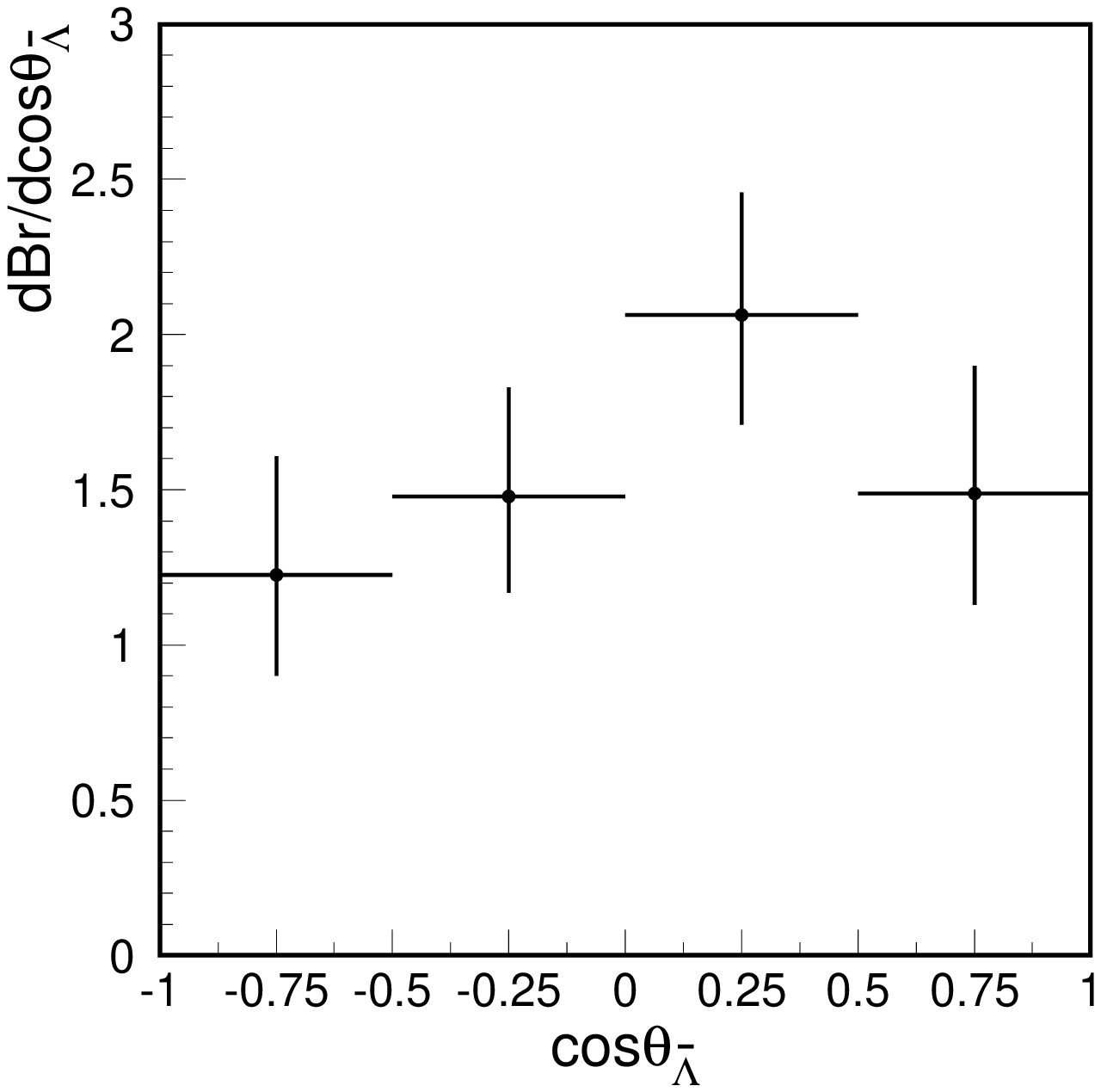}
  \includegraphics[width=0.33\textwidth]{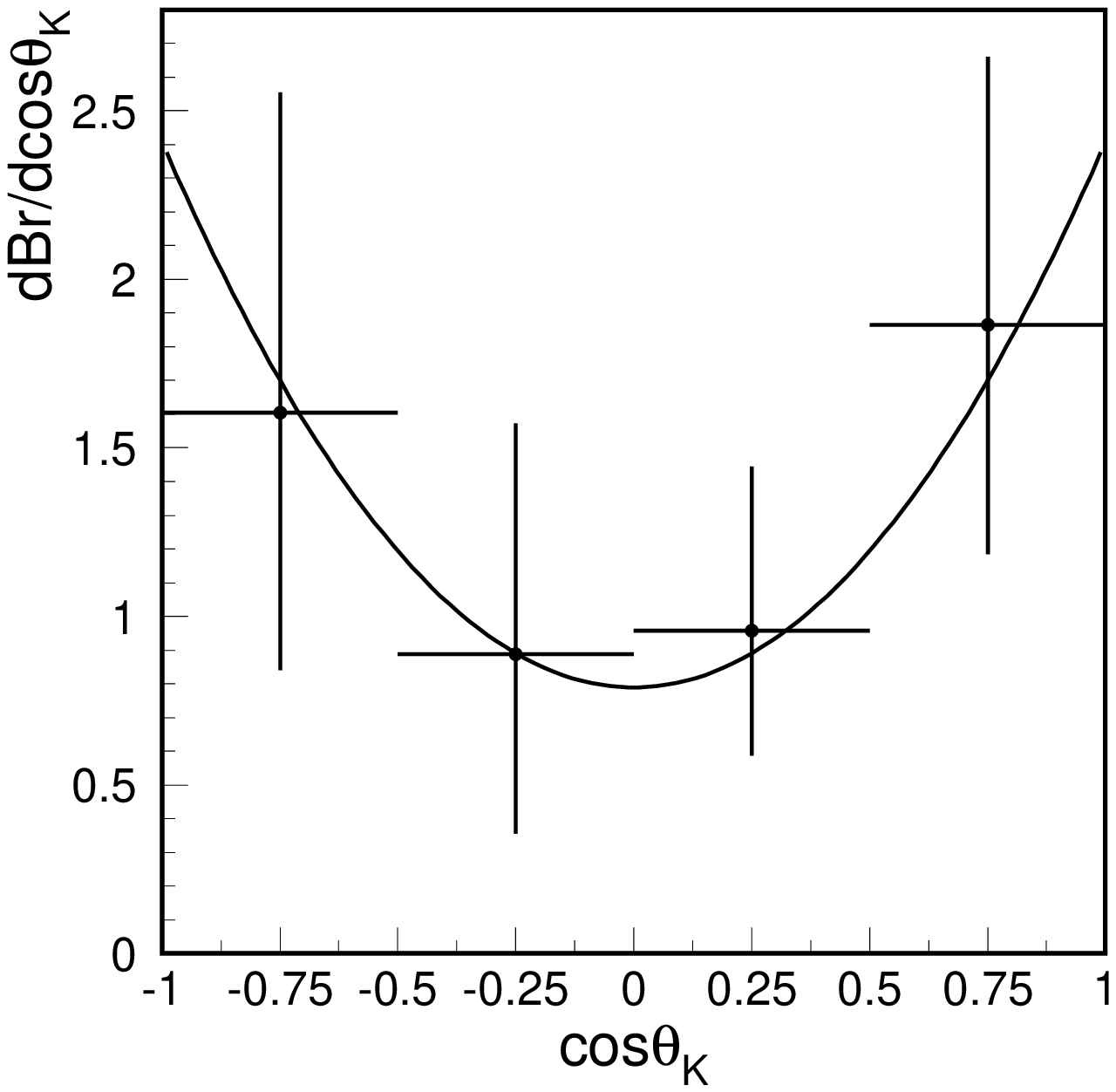}\\
 \vskip -0.4cm
 \caption{(a) Differential $\mathcal{B}(\bp \to \llk)$ {\it vs.} $\cos \theta_{\bar{\Lambda}}$
 in the $\LL$ rest frame. 
 (b) Differential $\mathcal{B}(\bz \to \llkstz)$ {\it vs.} $\cos \theta_K$  in the $K^{*0}$ system. 
 The solid curve represents the fitting result.
}
\label{fg:llk-costhl}
\end{center}
\end{figure}

Using 657 $ \times 10^6 B\bar{B}$ events, low mass $\mll$ enhancements near
threshold for both the $\llkz$ and $\llkstz$ modes are still observed, 
with 12.4$\sigma$ and 9.3$\sigma$ significance, respectively.
The branching fraction of $\bp \to \llk$ mode superseding the previous measurement~\cite{YJLee} is updated, 
and upper limits on the modes $\bp \to \llkst$ and $\bp \to \llpi$ in the region of $\mll < 2.85$ GeV/$c^2$ are set.
A related search for $B^0 \to \LL \bar{D}^0$ yields a branching fraction of
$\mathcal{B}(B^0 \to \LL \bar{D}^0) = (1.05^{+0.57}_{-0.44} \pm 0.14) \times 10^{-5}$ with 3.4$\sigma$. 
This may be compared with the large, $\sim 10^{-4}$, branching fraction observed for $B^0 \to \pp \bar{D}^0$. 
The $\mll$ enhancements near threshold and related angular distributions for the observed modes are also reported.
As what Figure~\ref{fg:llk-costhl} shows, 
the forward peaking distribution in $\bp \to \llk$ (as what $\bp \to \ppk$ decays do~\cite{JTWei}) is not observed,
and the fully polarized result in the helicity of $K^{*0}$ in $\bz \to \llkstz$ (as what 
$\bz \to \pp K^{*0}$ decays have~\cite{JHChen}) are not obtained.
All the details are summarized in Table~\ref{summation-table}.
The small value of $\mathcal{B} (\bp \to \llpi)$,
the large value of $\mathcal{B} (\bz \to \llkz)$,
and the absence of a peaking feature in the $\cos\theta_{\bar{\Lambda}}$ distribution
for $\bp \to \llk$ indicate that the
underlying dynamics of $B \to \LL h$ are quite different from those of $B \to \pp h$. 
These results also imply that the $\bar{s}$ quark from $\bar{b} \to \bar{s}$ penguin diagram does not necessarily 
hadronize to form a $K^+$;
the probability of forming a $\bar{\Lambda}$ is not negligible.
In addition, because $\mathcal{B} (\bz \to \LL \bar{D^0} (\bp \to \llpi)$)
is much  smaller than $\mathcal{B} (\bz \to \pp \bar{D^0} (\bp \to \pppi$)),
it appears that diquark pair popping out from the vacuum for
$us - \bar{u}\bar{s}$  ($sd - \bar{s}\bar{d}$) is suppressed compared to
$uu -\bar{u}\bar{u}$ ($ud - \bar{u}\bar{d}$).

\vspace{-1.0cm}
\begin{table}[hbp]
 \small
 \caption{Summary of measurements of $\mathcal{B}(B \to \LL h)$ with the corresponding significance $S$.}
 \begin{center}
  \begin{tabular}{|c|cc|c|cc|}
   \hline
   \multicolumn{3}{|c|}{Charmless branching fractions.} & \multicolumn{3}{c|}{Results in the region of $\mll < 2.85$ GeV/$c^2$.}\\
   \hline
   Mode               & $\mathcal B(10^{-6})$              & $S$          & 
   Mode               & $\mathcal B(10^{-6})$              & $S$          \\
   \hline
   $\bz \to \llkz$    & $4.76_{-0.68}^{+0.84} \pm 0.61$    & 12.5$\sigma$ &
   $\bp \to \llpi$    & $< 0.94$ at 90\% C.L.              & 2.5$\sigma$  \\
   $\bz \to \llkstz$  & $2.46_{-0.72}^{+0.87} \pm 0.34$    & 9.0$\sigma$  & 
   \multirow{2}{*}
   {$\bp \to \llkst$} & $< 4.98$ at 90\% C.L.              & 3.7$\sigma$  \\
   $\bp \to \llk$     & $3.38_{-0.36}^{+0.41} \pm 0.41$    & 16.5$\sigma$ & 
                      & $(2.19_{-0.88}^{+1.13} \pm 0.33)$  &              \\
   \hline
  \end{tabular}
 \end{center}
 \label{summation-table}
\end{table}

\section{Charmed baryonic \textit{B} decays}
\subsection{Study of intermediate two-body decays in $\bar{B}^0 \to \Sigma_c(2455)^0 \bar{p} \pi^+$}
In this analysis~\cite{HOKim}, a detailed study of the intermediate three-body decay 
$\bar{B}^0 \to \Sigma_c(2455)^0 \bar{p} \pi^+$ observed in the previous analysis of 
$\bar{B}^0 \to \Lambda_c^+ \bar{p} \pi^+ \pi^-$~\cite{KSPark} is performed, based on a data sample of $388 \times 10^6$ 
$B\bar{B}$ events, corresponding to $357 fb^{-1}$ accumulated at the $\Upsilon(4S)$ resonance with the Belle
detector. 
Hereafter, $\Sigma_c(2455)$ is denoted as $\Sigma_c$ in this analysis.

A broad $\bar{p} \pi^+$ mass structure near 1.5 GeV/$c^2$, denoted as $\bar{N}^0$, 
a uniform $cos\theta_{\bar{p}}$ distribution with a sharp forward peak , 
and a low mass $\Sigma_c^0 \pi^+$ enhancement, denoted as $X_{\Sigma_c^0 \pi^+}^+$, are observed.
Here, $cos\theta_p$ is the cosine of the angle between the $\bar{p}$ momentum and the direction
opposite to the $B$ momentum in the $\bar{p} \pi^+$ rest frame.
In order to describe the $\bar{p} \pi^+$ mass structure, which is not explained by 
a simple phase space non-resonant $\bar{B}^0 \to \Sigma_c^0 \bar{p} \pi^+$ decay, 
an intermediate two-body decay $\bar{p} \to \Sigma_c^0 \bar{N}^0$ with a resonant 
state $\bar{N}^0 \to \bar{p} \pi^+$ is considered. 
However, forward $cos\theta_p$ peak and the $\Sigma_c^0 \pi^+$ low mass structure 
still cannot be reproduced with these two modes only. 
Therefore, one additional mode $\bar{B}^0 \to X_{\Sigma_c^0 \pi^+}^+ \bar{p}$ is 
introduced to account for the observed features. 
As the low $M_{\Sigma_c^0 \pi^+}$ structure is close to threshold, 
it produces a forward peak in the $cos\theta_p$ distribution. 
Because a good candidate to interpret this broad structure
are not found within known $\Sigma_c^0 \pi^+$ resonant states,
a threshold mass enhancement is assumed.
Therefore, to explain these structures, contributions from an intermediate two-body decay 
$\bar{B}^0 \to \Sigma_c^0 \bar{N}^0$, non-resonant three-body decay $\bar{B}^0 \to \Sigma_c^0 \bar{p} \pi^+$ 
and a low mass structure near threshold $\bar{B}^0 \to X_{\Sigma_c^0 \pi^+}^+ \bar{p}$ are needed.
A simultaneous fit to the $M(\bar{p} \pi^+)$ and $cos\theta_{\bar{p}}$ distributions with those 
three modes is performed, and the yield and the relativistic Breit-Wigner parameters of the 
$\bar{N}^0$ state for $\bar{B}^0 \to \Sigma_c^0 \bar{N}^0$ are determined. 
The product of branching fractions is measured to be $\mathcal{B}(\bar{B}^0 \to \Sigma_c^0 \bar{N}^0) 
\times \mathcal{B}(\bar{N}^0 \to \bar{p} \pi^+) = (0.80 \pm 0.15(stat.) \pm 0.14(syst.) \pm 0.21) \times 10^{-4}$,
where the last error is due to the uncertainty in $\mathcal{B}(\Lambda_c^+ \to p K^- \pi+)$. 
The significance of the signal is 6.1$\sigma$ including systematics.
The fitted mass and width are consistent with the presence of an intermediate baryonic resonance $\bar{N}^0$, 
where $\bar{N}^0$ is the $\bar{N}(1440)^0 P_{11}$ or $\bar{N}(1535)^0 S_{11}$ state, or an admixture of the two states.

\subsection{Measurement of $\mathcal{B}(\bar{B}^0 \to \Lambda_c^+ \bar{p})$ and 
$\mathcal{B}(B^- \to \Lambda_c^+ \bar{p} \pi^-)$ and studies of $\Lambda^+ \pi^-$ resonances}
In this research~\cite{BAubert1}, the decays $\bar{B}^0 \to \Lambda_c^+ \bar{p}$ and 
$B^- \to \Lambda_c^+ \bar{p} \pi^-$ are investigated.
Baryon production in $B$ decays is studied 
by comparing the two-body $(\bar{B}^0 \to \Lambda_c^+ \bar{p})$ and 
three-body $(B^- \to \Lambda_c^+ \bar{p} \pi^-)$ decay rates directly. 
The dynamics of the baryon-antibaryon $(\Lambda_c^+ \bar{p})$ system in the
three-body decay provide insight into baryon production mechanisms. 
Additionally, the $B^- \to \Lambda_c^+ \bar{p} \pi^-$ system is a
laboratory for studying excited baryon states and is used to
measure the spin of the $\Sigma_c(2455)^0$. 
The measurements presented in this study are based on
$383 \times 10^6 ~\Upsilon(4S) \to B\bar{B}$ decays collected with the BABAR
detector at the PEP-II asymmetric-energy $e^+ e^-$ storage rings at the Stanford Linear Accelerator Center. 
The center-of-mass energy $\sqrt{s}$ corresponds to the mass of the $\Upsilon(4S)$ resonance at the interaction point.

The measured branching fractions and comparisons with the previous values~\cite{NGabyshev1,NGabyshev2} 
are shown in the table~\ref{tab:BF32}.
The ratio is $\mathcal{B}(B^- \to \Lambda_c^+ \bar{p} \pi^-)/\mathcal{B}(\bar{B}^0 \to \Lambda_c^+ \bar{p}) = 15.4 \pm 1.8 \pm 0.3$. 
For searches for excited charm baryon states in the $B^- \to \Lambda_c^+ \bar{p} \pi^-$ process, 
the resonant decays $B^- \to \Sigma_c(2455)^0 \bar{p}$ and $B^- \to \Sigma_c(2800) \bar{p}$ are observed, 
but the evidence for $B^- \to \Sigma_c(2520)^0 \bar{p}$ is not found.
{\small
\begin{equation}
 \begin{split}
  \frac{\mathcal{B}(B^- \to \Sigma_c(2455)^0 \bar{p})}{\mathcal{B}(B^- \to \Lambda_c^+ \bar{p} \pi^-)} &= (12.3 \pm 1.2 \pm 0.8) \times 10^{-2} \\
  \frac{\mathcal{B}(B^- \to \Sigma_c(2800)^0 \bar{p})}{\mathcal{B}(B^- \to \Lambda_c^+ \bar{p} \pi^-)} &= (11.7 \pm 2.3 \pm 2.4) \times 10^{-2} \\
  \frac{\mathcal{B}(B^- \to \Sigma_c(2520)^0 \bar{p})}{\mathcal{B}(B^- \to \Lambda_c^+ \bar{p} \pi^-)} &< 0.9 \times 10^{-2} ~(90\% ~C.L.)
 \end{split}
 \label{eq:RatioOfBF23}
\end{equation}
}

\vspace{-0.2cm}
This is the first observation of the decay $B^- \to \Sigma_c(2800) \bar{p}$; 
however, the mass of the observed excited $\Sigma_c^0$ state is $(2846 \pm 8 \pm 10)$ MeV/$c^2$, 
which is somewhat inconsistent with PDG value~\cite{CAmsler}. 
Finally, the spin of the $\Sigma_c(2455)^0$ baryon is measured to be 1/2 as predicted by the quark model.
This is the first measurement of the spin of this state.

\begin{table}[htp]
 \small
 \caption{Measurements of branching fractions $\mathcal{B}(\bar{B}^0 \to \Lambda_c^+ \bar{p})$ and 
 $\mathcal{B}(B^- \to \Lambda_c^+ \bar{p} \pi^-)$ at Babar and Belle. The uncertainty 
 on $\mathcal{B}$ are statistical, systematic, and the uncertainty due to the uncertainty in 
 $\mathcal{B}(\Lambda_c^+ \to p K^- \pi^+)$.}
 \begin{center}
  \begin{tabular}{|l|ll|}
   \hline
   Decay Mode & Babar $(10^{-5})$ & Belle $(10^{-5})$\\
   \hline
   $\mathcal{B}(\bar{B}^0 \to \Lambda_c^+ \bar{p})$ 
              & $1.89 \pm 0.21 \pm 0.06 \pm 0.49$ & $2.19~^{+0.56}_{-0.49} \hspace{0.8pt}\pm 0.32 \pm 0.57$\\
   $\mathcal{B}(B^- \to \Lambda_c^+ \bar{p} \pi^-)$
              & $33.8 \pm 1.2~ \hspace{1.8pt}\pm 1.2~ \hspace{1.8pt}\pm 8.8~~$    & $20.1 \pm 1.5 \pm 2.0~ \hspace{1.8pt}\pm 5.2~~$\\
   \hline
  \end{tabular}
 \end{center}
 \label{tab:BF32}
\end{table}


\subsection{Study of $\bar{B} \to \Xi_c \bar{\Lambda}_c^-$ and $\bar{B} \to \LcLc \bar{K}$ decays}
This study~\cite{BAubert2} reports measurements of $B$-meson decays into two- and three-body final states containing 
two charmed baryons using a sample of $230 \times 10^6 ~\Upsilon(4S) \to B\bar{B}$ decays. 
Measurements of the branching fraction of the decays $B^- \to \LcLc K^-$, $B^- \to \XcLcm$, 
$\bar{B}^0 \to \Xi_c^+ \bar{\Lambda}_c^-$, and $\bar{B}^0 \to \LcLc \bar{K}^0$ are presented, and three-body 
decays for the possible presence of intermediate resonances are investigated. 
The data were collected with the BABAR detector and 
represent an integrated luminosity of approximately $210 ~fb^{-1}$ collected at the $\Upsilon(4S)$ resonance.

The measured branching fractions obtained from modes with significant signals are 
$\mathcal{B}(B^- \to \Xi_c^0 \bar{\Lambda}_c^-) \times (\Xi_c^0 \to \Xi^- \pi^+) = (2.08 \pm 0.65 \pm 0.29 \pm 0.54) \times 10^{-5}$ and
$\mathcal{B}(B^- \to \LcLc K^-) = (1.14 \pm 0.15 \pm 0.17 \pm 0.60) \times 10^{-3}$, 
where the uncertainties are statistical, systematic, 
and from the branching fraction $\mathcal{B}(\Lambda_c^+ \to p K^- \pi^+)$, respectively. 
On two other modes, upper limits at the 90\% confidence level are set:
$\mathcal{B}(\bar{B}^0 \to \Xi_c^+ \bar{\Lambda}_c^-) \times \mathcal{B}(\Xi_c^+ \to \Xi^- \pi^+ \pi^+) <  \times 10^{-5}$ and 
$\mathcal{B}(\bar{B}^0 \to \LcLc \bar{K})^0 < 1.5 \times 10^{-3}$. 
As table~\ref{tab:Xic-Lambdac} and \ref{tab:Lambdac-Lambdac} show, 
the measurements are consistent with the previous values within uncertainties~\cite{RChistov,NGabyshev3}. 
The branching fraction of $B^- \to \LcLc K^-$ is found to be comparable to the $\mathcal{O}(10^{-3})$ branching fraction predicted for 
two-body decays to a pair of charmed baryons. 
In the $\Lambda_c^+ K^-$ mass distribution of the decay $B^- \to \LcLc K^-$, 
a structure centered at an invariant mass of 2.93~GeV/$c^2$ is observed.
The data in the Dalitz plot and
two-body mass projections are inconsistent with a phase space
distribution and suggest the presence of a $\Xi_c^0$
resonance in the decay.

\vspace{-9pt}
\begin{table}[htb]
 \small
 \caption{Measurements of $\mathcal{B}(B^- \to \XcLcm)$ and $\mathcal{B}(\bar{B}^0  \to \Xi_c^+ \bar{\Lambda}_c^-)$ 
 at Babar and Belle with the corresponding significance $S$. The uncertainties are statistical, systematic, and the 
 uncertainty of $\mathcal{B}(\bar{\Lambda}_c^- \to \bar{p} K^+ \pi^-)$, respectively.}
 \begin{center}
  \begin{tabular}{|l|cc|cc|}
   \hline
   Decay Mode & Babar $(10^{-5})$ & $S$ & Belle $(10^{-5}$) & $S$\\
   \hline
   $\mathcal{B}(B^- \to \XcLcm)$ 
              & $2.08 \pm 0.65 \pm 0.29 \pm 0.54$   & 6.4$\sigma$
              & $4.8^{+1.0}_{-0.9} \pm 1.1 \pm 1.2$ & 8.7$\sigma$\\
   \hspace{2pt} $\mathcal{B}(B^- \to \XcLcm) \times \mathcal{B}(\Xi_c^0 \to \Xi^- \pi^+)$
              & $2.51 \pm 0.89 \pm 0.29 \pm 0.65$   & 6.1$\sigma$
              & $5.6^{+1.9}_{-1.5} \pm 1.1 \pm 1.5$ & 6.8$\sigma$\\
   \hspace{2pt} $\mathcal{B}(B^- \to \XcLcm) \times \mathcal{B}(\Xi_c^0 \to \Lambda K^- \pi^+)$
              & $1.70 \pm 0.93 \pm 0.30 \pm 0.44$   & 2.1$\sigma$ 
              & $4.0^{+1.1}_{-0.9} \pm 0.9 \pm 1.0$ & 5.9$\sigma$\\
   $\mathcal{B}(\bar{B}^0 \to \Xi_c^+ \bar{\Lambda}_c^-) \times \mathcal{B}(\Xi_c^+ \to \Xi^- \pi^+ \pi^+)$
              & $1.50 \pm 1.07 \pm 0.20 \pm 0.39$   & 1.8$\sigma$ 
              & $9.3^{+3.7}_{-2.8} \pm 1.9 \pm 2.4$ & 3.8$\sigma$\\
   \hline
  \end{tabular}
 \end{center}
 \label{tab:Xic-Lambdac}
\end{table}

\vspace{-24pt}
\begin{table}[htb]
 \small
 \caption{Measurements of $\mathcal{B}(B^- \to \LcLc K^-)$ and $\mathcal{B}  (\bar{B}^0 \to \LcLc \bar{K}^0)$ 
 at Babar and Belle with the corresponding significance $S$. The uncertainties are statistical, systematic, 
 and the uncertainty of $\mathcal{B}(\Lambda_c^+ \to p K^- \pi^+)$, respectively.}
 \begin{center}
  \begin{tabular}{|l|cc|cc|}
   \hline
   Decay Mode & Babar $(10^{-4})$ & $S$ & Belle $(10^{-4})$ & $S$\\
   \hline
   $\mathcal{B}(B^- \to \LcLc K^-)$
              & $11.4 \pm 1.5 \pm 1.7 \pm 6.0$      &  9.6$\sigma$
              & $6.5^{+1.0}_{-0.9} \pm 0.8 \pm 3.4$ & 15.4$\sigma$\\
   $\mathcal{B}(\bar{B}^0 \to \LcLc \bar{K}^0)$ 
              & $3.8 \pm 3.1 \pm 0.5 \pm 2.0$       &  1.4$\sigma$
              & $7.9^{+2.9}_{-2.3} \pm 1.2 \pm 4.2$ &  6.6$\sigma$ \\
   \hline
  \end{tabular}
 \end{center}
 \label{tab:Lambdac-Lambdac}
\end{table}

\vspace{-10pt}{
\end{document}